\documentclass[a4paper,fleqn,usenatbib]{mnras}

\usepackage[T1]{fontenc}
\usepackage{ae,aecompl,color}

\usepackage{graphicx}	
\usepackage{amsmath}	
\usepackage{amssymb}	

\title[Feedback-driven star formation in NGC 1893]{Kinematic evidence 
for feedback-driven star formation in NGC 1893}

\author[B. Lim et al.]{Beomdu Lim,$^{1,2}$\thanks{Corresponding author, E-mail:blim@uliege.be}
  Hwankyung Sung,$^2$
  Michael S. Bessell,$^3$
  Sangwoo Lee,$^4$
  \newauthor Jae Joon Lee$^5$, Heeyoung Oh$^5$, Narae Hwang$^5$, 
  Byeong-Gon Park$^{5,6}$, Hyeonoh Hur$^7$, 
  \newauthor Kyeongsoo Hong$^5$, and 
  Sunkyung Park$^8$ 
\\
$^1$Space sciences, Technologies and Astrophysics Research (STAR) Institute, 
                  Universit\'e de Li\'ege, Quartier Agora, All\'ee du Ao\^ut 19c,\\
                  B\^at. B5C, 4000, Li\'ege, Belgium\\
  $^2$Department of Astronomy and Space Science, Sejong University, 209 Neungdong-ro, 
              Gwangjin-gu, Seoul 05006, Republic of Korea\\
  $^3$Research School of Astronomy and Astrophysics, The Australian National University, 
              Canberra, ACT 2611, Australia\\
  $^4$SELab, Inc., 8 Nonhyeon-ro 150-gil, Gangnam-gu, Seoul 06049, Republic of Korea\\
  $^5$Korea Astronomy and Space Science Institute, 776 Daedeokdae-ro, Yuseong-gu, Daejeon 305-348, Korea \\
  $^6$Astronomy and Space Science Major, University of Science and Technology, 217 
             Gajeong-ro, Yuseong-gu, Daejeon 34113, Republic of Korea\\
  $^7$Daegu National Science Museum, 20, Techno-daero 6-gil, Yuga-myeon, Dalseong-gun, Daegu 
             43023, Republic of Korea\\
  $^8$School of Space Research, Kyung Hee University
            1732, Deogyeong-daero, Giheung-gu, Yongin-si, Gyeonggi-do, 17104, Republic of Korea
}

\date{Accepted XXX. Received YYY; in original form ZZZ}

\pubyear{2018}

\begin{document}
\label{firstpage}
\pagerange{\pageref{firstpage}--\pageref{lastpage}}
\maketitle

\begin{abstract}
OB associations are the prevailing star forming sites in the Galaxy. Up to now, 
the process of how OB associations were formed remained a mystery. A possible 
process is self-regulating star formation driven by feedback from massive stars. 
However, although a number of observational studies uncovered various signposts 
of feedback-driven star formation, the effectiveness of such feedback has been 
questioned. Stellar and gas kinematics is a promising tool to capture the relative 
motion of newborn stars and gas away from ionizing sources. We present 
high-resolution spectroscopy of stars and gas in the young open cluster NGC 1893. Our 
findings show that newborn stars and the tadpole nebula Sim 130 are moving away from 
the central cluster containing two O-type stars, and that the timescale of sequential 
star formation is about 1 Myr within a 9 parsec distance. The newborn stars 
formed by feedback from massive stars account for at least 18 per cent of the total stellar 
population in the cluster, suggesting that this process can play an important role in the 
formation of OB associations. These results support the self-regulating star formation 
model.
\end{abstract}

\begin{keywords}
stars: formation -- stars: kinematics and dynamics -- HII regions -- 
ISM: kinematics and dynamics -- open clusters and associations: individual (NGC 1893) 
\end{keywords}



\section{Introduction}
A giant molecular cloud can form a few star-forming complexes with scales of tens of parsecs. 
The resultant stellar system is a so-called OB association consisting of 
loose groups of massive OB stars \citep{A47}. OB associations are considered as 
the prime sites of star formation and factories of field stars in the Galaxy \citep{BPS07,BBH99,MS78}. 
However, their formation process is so far not well established. 
One reason is the large extent of these objects, and the other is 
the difficulty of identifying low-mass members from amongst the contaminating 
foreground and background stars. With the advent of high-sensitivity instruments 
at infrared and X-ray wavelengths, a large number of low-mass members have now been 
identified enabling the mystery of the internal structure and formation process of OB-associations
to start to be understood.

\begin{figure*}
   \centering
   \includegraphics[width=16cm]{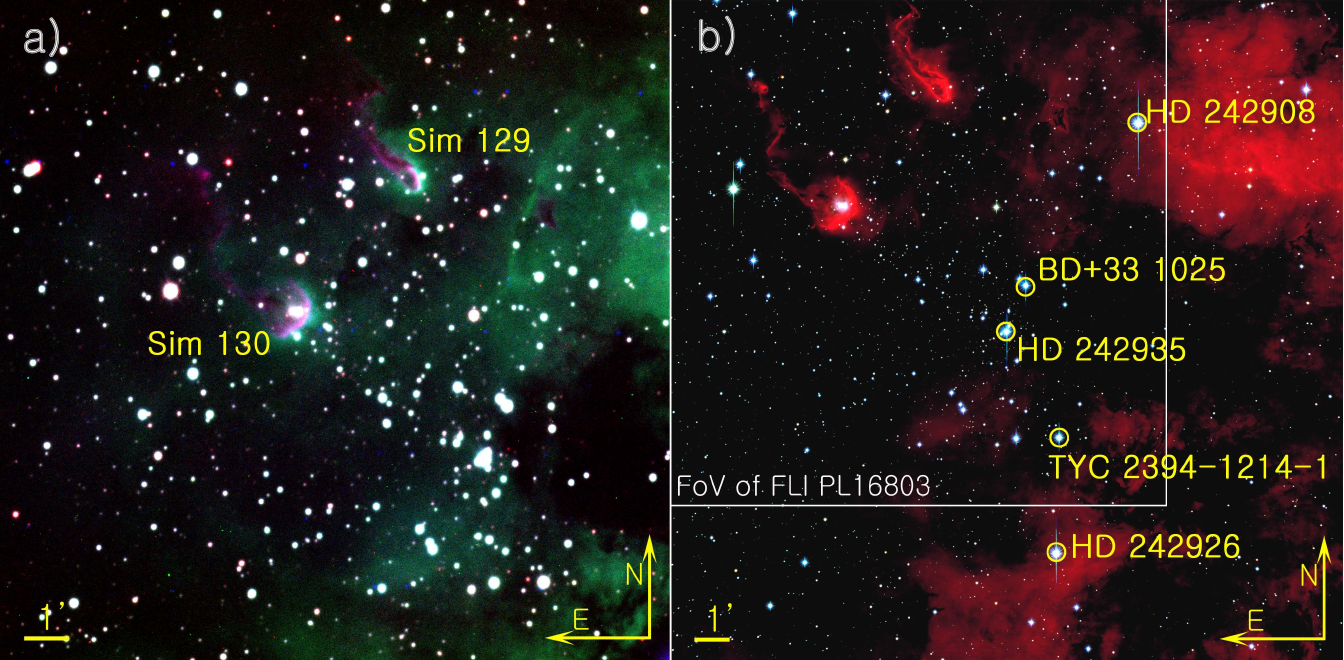}
   \caption{Colour composite images of NGC 1893 ({\bf a.} blue : H$\beta$, green : 
[O {\scriptsize \textsc{III}}] $\lambda$5007, and red : [S {\scriptsize \textsc{II}}] 
$\lambda$6712, {\bf b.} blue : $B$, green: $V$, and red : H$\alpha$). The positions 
of the tadpole nebulae and O-type stars are labelled on each image. The field of view 
is about $15\farcm8 \times 15\farcm8$ for the FLI PL16803 CCD camera 
({\bf a}) and $18^{\prime} \times 18^{\prime}$ for SNUCam ({\bf b}). An overlap 
region between the fields of view of the CCD cameras is outlined by a white solid
 line.}
              \label{fig1}
\end{figure*}

OB associations, in general, comprise highly concentrated core clusters 
and subgroups of low-stellar density \citep{B64,KAG08,SBC17}. This substructure 
may reflect their formation process. There are three models addressing the origin of OB associations. 
From N-body simulations, \citet{KAH01} proposed that about 30 per cent of stars in an embedded 
cluster can survive as a bound cluster for about a hundred million years while the other stars 
become a distributed population after the gas is expelled. According to this model, OB associations 
originated from the dynamical evolution of embedded clusters. On the other hand, \citet{CBZB05} claimed 
that a turbulent unbound giant molecular cloud can form several star clusters containing 
OB stars on their crossing timescale. In the model, these clusters are expanding 
having inherited the kinematics of their unbound natal cloud. A clustering of such 
star clusters independently formed from their natal cloud can have a large spatial distribution 
over several parsecs as seen in OB associations. However, neither of these models can explain the 
substructure within OB associations that is related to their star formation history. 

An alternative model is self-regulating star formation driven by feedback from massive 
O-type stars \citep{EL77}. Massive stars can sculpt their natal cloud, creating pillar-like 
structures and bright-rimmed clouds at the border of H {\scriptsize \textsc{II}} regions, 
as well as ionized gas bubbles \citep{KAG08}. A number of sophisticated computer 
simulations have successfully reproduced these gas structures in ultraviolet radiation 
fields \citep{GNB09,GNW09,DEB12,HHA12}. Subgroups of young stellar objects are 
often found in the vicinity of such gas structures, and moreover, an age sequence 
among them has been reported from observations \citep[etc.]{FHS02,KAG08,ZRM10,LSK14}. 
In addition, the shape of such subgroups appears to be elongated towards the
ionizing sources \citep[eg.][]{GFG07,CPO09}. These findings are believed 
to be observational evidence for positive feedback from massive stars. However, caveats 
from simulations are that all these signs are not necessarily indicative of feedback-driven 
star formation, and that observationally, stars formed by feedback would be 
indistinguishable from those spontaneously formed \citep{DEB13,DHB15}.

NGC 1893, a part of the Auriga OB2 association, is ideally suited to validate the 
self-regulating star formation model because this cluster shows typical signs of 
feedback-driven star formation. There are a total of five O-type stars, and a number 
of young stellar objects have been identified in the vicinity of the tadpole nebulae 
Sim 129 and 130 \citep{MSB07,NMIB07}. The spatial distribution of the 
members also shows a highly elongated shape from the cluster centre to the tadpole 
nebulae \citep{KGF15}, and an age difference between them has been found 
\citep{SPO07,MSB07,PSC13}. 

In \citet{LSK14}, we performed a comprehensive photometric study of NGC 
1893. A total of 906 stars were selected as members of NGC 1893 using a colour 
and magnitude cut method, H$\alpha$ photometry, and the published catalogues of 
young stellar objects and X-ray sources \citep{CMP08,CMP12}. Based on this membership 
list, the fundamental parameters of NGC 1893, such as reddening ($\langle E(B-V)\rangle 
= 0.56\pm 0.08$), distance ($3.5\pm0.3$ kiloparsecs), age (1.5 Myr), and the initial 
mass function ($\Gamma = -1.3 \pm 0.1$), were derived by analyzing various 
photometric diagrams. An age difference of 0.5 Myr was found from comparison of the 
age distributions of pre-main sequence (PMS) stars in the cluster centre and Sim 130, 
and we claimed that the age difference is a result of sequential star formation by feedback 
from the O-type stars in the cluster centre. Measuring the radial velocities (RVs) of the stars 
and gas would allow us to kinematically identify newborn stars formed by feedback from 
a coeval population of ionizing sources and to infer their contribution in the formation 
of an OB association.

In the present work, we present the high-resolution spectroscopic observations of 
member stars and ionized gas in NGC 1893. The kinematic substructure of the cluster 
provides further evidence for a new generation of stars formed by feedback from 
the first generation of massive stars. The observations and data reduction are described 
in Section 2. The measurement of RVs is addressed in Section 3. The kinematic substructure 
and age trend within the cluster are investigated with various diagrams
in Section 4 and 5, respectively. Multiple sets of Monte-Carlo simulations have been conducted, and the 
implications are discussed in Section 6. Finally, we present a summary and conclusions 
of this study in Section 7.

\begin{table*}
\caption{A summary of observations}\scriptsize
\label{tab1}
\centering
\begin{tabular}{ccccccc}\hline
UT date & Target & Telescope & Instrument &  Filters & Exposure time & binning \\
(YY-MM-DD) &   &  & & & (seconds)& \\
\hline
Imaging observations\\
\hline  
2009-01-19 & NGC 1893      & 1.5-m AZT-22 & SNUCam                                        &  $U$ & 30, 600 & $1\times1$\\  
                   &                        &                        &                                                         &  $B$  & 20, 600 & $1\times1$\\
                   &                        &                        &  $18^{\prime}\times18^{\prime}$     &   $V$  & 10, 300 & $1\times1$ \\
                   &                        &                        &                                                         &   $I$   & 5, 120 & $1\times1$\\ 
                   &                        &                        &                                                         &  H$\alpha$ & 60, 600 & $1\times1$\\
\hline
2016-11-28 &NGC 1893       & 1-m at DOAO    & FLI PL16803 &   H$\beta$ & 1200 & $2\times2$ \\
                   &                       &                        &  $15\farcm8\times15\farcm8$ & [O {\scriptsize \textsc{III}}] $\lambda5007$ & 900 & $2\times2$\\
                   &                       &                        &                                   &     [S {\scriptsize \textsc{II}}] $\lambda6712$ & 600 & $2\times2$ \\
                   
\hline
Spectroscopic observations\\
\hline
2016-01-26& 14 stars & 6.5-m MMT & Hectochelle &   RV31 (5150\AA  \ -- 5300\AA) & $3\times300$ & $2\times2$\\
2016-01-26& 14 stars & &                     &   RV31 (5150\AA  \ -- 5300\AA) & $3\times300$ & $2\times2$\\
2016-01-26&  48 stars& &                     &   RV31 (5150\AA  \ -- 5300\AA) & $3\times1200$ & $2\times2$\\
2016-01-27& 240 positions & &                     &   OB25 (6475\AA  \ -- 6630\AA) & $3\times900$ & $1\times1$\\
2016-01-27& 240 positions     & &                     &   OB25 (6475\AA  \ -- 6630\AA) & $3\times900$ & $1\times1$\\
2016-01-28& 50 stars & &                     &   RV31 (5150\AA  \ -- 5300\AA) & $3\times1200$ & $2\times2$\\
2016-01-31&    57 stars                   & &                     &   RV31 (5150\AA  \ -- 5300\AA) & $3\times2700$ & $2\times2$\\
\hline
2015-12-17&HD 242935  & 1.8-m at BOAO & BOES & 3800\AA \ -- 9000\AA & $3\times 1200$ & $1\times1$ \\
                  &BD +33 1025 &                           &           &                                    & $3\times 1200$ &                    \\
                  &HD 242908   &                           &           &                                    &$3\times 1200$  &                   \\
                  &TYC 2394-1500-1 &                  &            &                                    &$3\times 1200$  &                   \\
                  &TYC 2394-1214-1 &                   &           &                                    &$3\times1200$   &                    \\
2015-12-27&HD 242926  &                            &           &                                    &$3\times 1200$ &                      \\
2016-02-01&BD +33 1025 &                          &            &                                    &$3\times 1200$ &                      \\
2016-02-02&HD 242908   &                          &            &                                    &$3\times 1200$ &                      \\
2016-02-03&HD 242926  &                           &            &                                    &$3\times1200$ &                      \\
2016-02-04&HD 242935 &                            &            &                                    &$3\times 1200$ &                      \\
\hline
2016-03-02& Sim 130-01       & 2.7-m HJST & IGRINS      &   $H$ and $K$ bands (1.45$\mu$m -- 2.50$\mu$m) & $3\times300$ & $1\times1$ \\
                  & Sim 130-02       & &                    &                                                                                   & $2\times300$ &      \\
                  & Sim 130-03       & &                    &                                                                                   & $2\times300$ & \\
\hline
2016-11-18 &Sim 129-00       & 4.3-m DCT & IGRINS       &   $H$ and $K$ bands (1.45$\mu$m -- 2.50$\mu$m) & $2\times300$ & $1\times1$ \\
                   & Sim 129-01      & &                    &                                                                                   & $2\times300$ &                  \\
                   &Sim 129-02       & &                    &                                                                                   & $2\times300$ &  \\
                   & Sim 129-03      & &                    &                                                                                   & $2\times300$ &  \\
                   & Sim 129-04      & &                    &                                                                                   & $2\times300$ &  \\
                   &Sim 129-05       & &                    &                                                                                   & $2\times300$ &  \\
2016-11-19 &Sim 130-00       & &                   &                                                                                  & $2\times300$ & $1\times1$ \\
                   &Sim 130-04       & &                    &                                                                                   & $2\times300$ &  \\
                   &Sim 130-05       & &                    &                                                                                   & $2\times300$ &  \\
                   &Sim 130-06       & &                    &                                                                                   & $2\times300$ &  \\
                   &Sim 130-07       & &                    &                                                                                   & $2\times300$ &  \\
                   &Sim 130-08       & &                    &                                                                                   & $2\times300$ &  \\
                   &Sim 130-09       & &                    &                                                                                   & $2\times300$ &  \\
                   &Sim 130-10       & &                    &                                                                                   & $2\times300$ &  \\
 \hline                  
\end{tabular}
\end{table*}

\section{Observations and data reduction}
\subsection{Imaging observations} 
We observed a $15\farcm8\times15\farcm8$ region of NGC 1893 using the 
FLI PL16803 CCD camera on the 1-m telescope at Deokheung Optical Astronomy Observatory 
in Korea on 2016 November 28. The narrow band H$\beta$, [O {\scriptsize \textsc{III}}] 
$\lambda5007$, and [S {\scriptsize \textsc{II}}] $\lambda 6712$ filters were used to study 
the nebulosity of ionized gas across the cluster. Raw images were subtracted by a master 
bias image, and subsequently divided by master flat images for given bands. A colour 
composite image was created by combining the reduced images (see Fig. 1-a).   

SNUCam is a 4k$\times$4k CCD camera installed on the 1.5-m AZT-22 telescope 
($f/7.74$) at Maidanak Astronomical Observatory in Uzbekistan \citep{IKC10}. We 
observed a $18^{\prime}\times18^{\prime}$ region of NGC 1893 on 2009 January 19 
using this camera with the broad band $UBVI$ and narrow band H$\alpha$ filters. Image 
pre-processing and photometry of stars are described in our previous work in 
detail \citep{LSK14}. A colour composite image was created using $B$, $V$, and 
H$\alpha$ images (see Fig. 1-b).

Early-type stars (O- and B-type) have prominent blue $U-B$ colours, and 
their reddening and distance can easily be constrained from the ($U-B$, $B-V$) 
diagram and colour-magnitude diagrams, respectively. The member selection of these 
stars was made by using a colour and magnitude cut method (see \citealt{LSK14} for detail). 
PMS stars tend to have circumstellar discs and are very active objects. X-ray emission 
and some H$\alpha$ emission is related to magnetic activity in the young star while 
strong H$\alpha$ emission arises from the disc accretion and a warm circumstellar 
disc emits strongly at infrared wavelengths.  We identified 
H$\alpha$ emission star candidates from H$\alpha$ photometry. A list of PMS 
stars showing mid-infrared excess emission or X-ray emission were taken from 
the catalogues of young stellar objects \citep{CMP08,CMP12}. A total of 906 stars 
were selected as members of NGC 1893 \citep{LSK14}.

\subsection{Spectroscopic observations}
Queue scheduled observations of 183 target stars out of 906 cluster 
members were carried out on 2016 January 26, 27, 28, and 31 with the 
multi-object high-resolution spectrograph Hectochelle attached to the 
6.5-m telescope of the MMT observatory. The spectral resolving power 
of Hectochelle is about $R \sim 34000$. A total of 240 fibres can be 
simultaneously used to observe targets and sky background in a single 
exposure \citep{SFC11}. The observations of ionized gas were made 
with the order-separating filter OB 25 transmitting light in the wavelength 
range of 6475\AA \ -- 6630\AA \ in a $1\times1$ binning mode, while 
we obtained the spectra of stars in $2\times2$ binning mode to achieve 
a good signal-to-noise ratio using the RV 31 filter covering 5150\AA \ 
-- 5300 \AA. Calibration spectra, such as dome flat and ThAr lamp spectra, 
were also taken, just before and after the target exposure. 

All the frames were written in the multiple extension Flexible Image Transport 
System (FITS). We merged these mosaic images into a single FITS image after overscan 
correction with \textsc{IRAF/MSCRED} packages. One-dimensional (1D) spectra of 
targets were extracted using the {\it dofibers} task in the \textsc{IRAF/SPECRED} package. 
Odd- and even-numbered apertures were traced in the dome flat spectra using a spline 
function fitting method, respectively, and subsequently the 1D spectra of targets and 
sky were extracted along corresponding apertures. We also corrected for a variation in 
the pixel-to-pixel response from residuals of the dome flat divided by a high-order 
spline function. The solutions for the wavelength calibration were determined from ThAr 
lamp spectra, applied to the dome flat, targets, and sky spectra. The spectra of the targets 
and sky were flattened by eliminating the blaze function of the echelle spectra from 
the wavelength-corrected dome flat spectra. 

We created a master sky spectrum with a high signal-to-noise ratio by median 
combining the spectra from a few tens of fibres assigned to the blank sky. 
This master sky spectrum was used to subtract the contribution of airglow to the spectra 
of stars. The sky-subtracted spectra for the same target were combined into a single 
spectrum with a better signal-to-noise ratio and normalized by using continuum 
levels found from a cubic spline interpolation. On the other hand, we combined the 
spectra of the ionized gas for the same fibre position into a single spectrum.

Spectroscopic observations of six OB stars were carried out with the fibre-fed Bohyunsan 
Observatory Echelle Spectrograph \citep[BOES]{KHV07} attached to the 1.8-m telescope 
at Bohyunsan Optical Astronomy Observatory in Korea in December 2015 and February 2016. We 
acquired the spectra of the stars with a 300 $\mu$m fibre whose spectral resolution 
is about $R = 30000$. Calibration frames were obtained either at the end of each 
night or at the beginning of observations. Pre-processing and extraction of the 
spectra were conducted with the \textsc{IRAF/ECHELLE} package in almost the 
same way as the data reduction procedure of Hectochelle. We applied the wavelength 
calibration obtained from the ThAr lamp spectra to the extracted 1D target spectra, 
and then those target spectra were flattened and normalized.  

The Immersion GRating INfrared Spectrograph (IGRINS) simultaneously obtained the 
near-infrared $H$ and $K$ band spectra with a high spectral resolution of $R \sim 
45000$ \citep{YJB10,PJY14}. This spectrograph was operated with the 2.7-m Harlan
 J. Smith Telescope at McDonald Observatory of the University of Texas at Austin until 
the first half year of 2016. Pilot observations of three slit positions on the head of the 
tadpole nebula Sim 130 was made on 2016 March 2. In the fall of 2016, IGRINS was 
installed on the 4.3-m Discovery Channel Telescope at Lowell Observatory. We observed 
six slit positions on the head of Sim 129 and eight positions on that of Sim 130 on 2016 
November 18 and 19, respectively. In all the observations, a nod technique, ON (source) 
- OFF (sky) - ON (source), was applied to the observing sequence for removing the 
background flux. In order to eliminate a large number of telluric lines, the A0 main 
sequence stars 136 Tau, BD +27 716B, and HR 1558 were observed at almost the same 
air mass as the tadpole nebulae. 

We performed data reduction using the IGRINS Pipeline Package\footnote{The 
IGRINS Pipeline Package can be download at http://github.com/igrins/plp. 
doi:10.5281/zenodo.18579}. This pipeline contains basic procedures for data 
reduction, such as aperture extraction, the subtraction of background emission, bad 
pixel correction, and wavelength calibration. Subsequently, after correction for telluric 
lines using the 1D spectrum of an observed A0 main sequence star with the public tool 
\textsc{PLOTSPEC} \citep{KDO17}, we stitched together all the echelle spectra to make 
a two-dimensional relative flux calibrated long spectrum over the H and K bands. We 
summarized our observations in Table~\ref{tab1}.

\begin{figure}
   \centering
   \includegraphics[width=8cm]{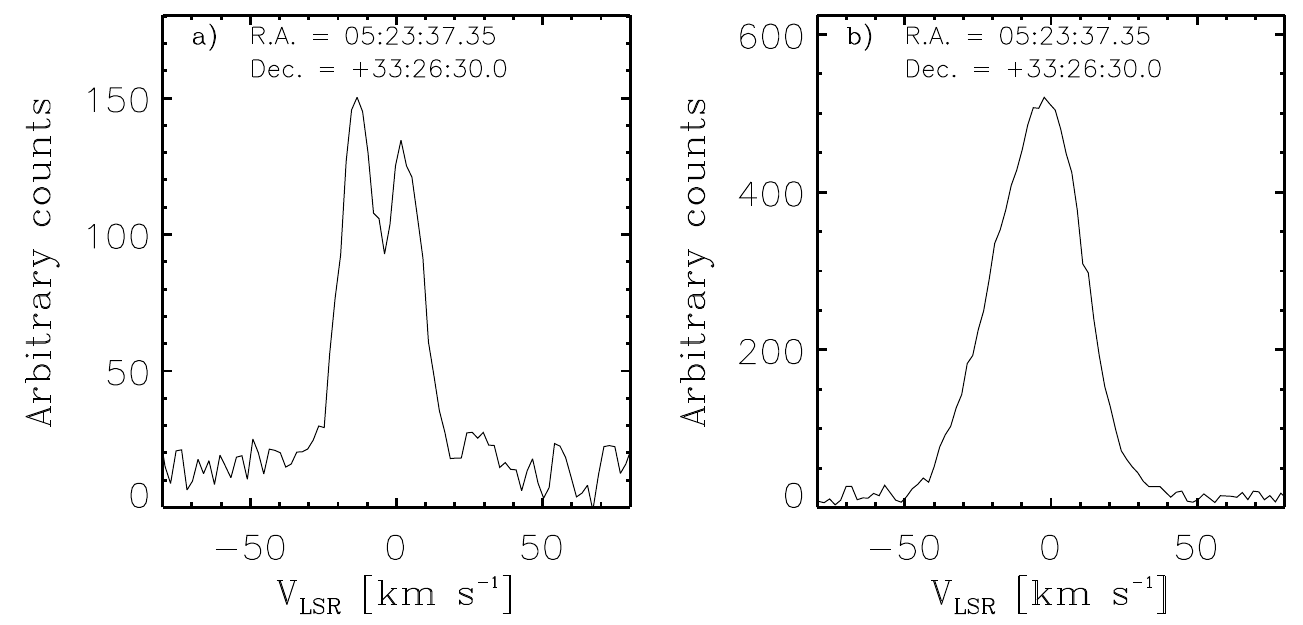}
      \caption{[N {\scriptsize \textsc{II}}] $\lambda$6584 ({\bf a}) and H$\alpha$ ({\bf b}) 
emission lines from the spectrum for a given fibre position. The forbidden line [N {\scriptsize 
\textsc{II}}] $\lambda$6584 shows a few components with different radial velocities in the line 
of sight due to absence of absorption column for its spontaneous transition, whereas the 
H$\alpha$ line exhibits a simple line profile with a strong single peak at near the systemic 
velocity of NGC 1893. The processes of reabsorption and emission by prevailing 
hydrogen atoms with the quantum number $n = 2$ in the intracluster medium are responsible 
for the formation of the observed H$\alpha$ line profile. Therefore, [N {\scriptsize \textsc{II}}] 
$\lambda$6584 line is a more useful tool to probe the velocity field of ionized gas rather 
than H$\alpha$ emission line.}
         \label{fig2}
   \end{figure}

\section{Radial velocities of gas and stars}
The forbidden line [N {\scriptsize \textsc{II}}] $\lambda6584$ was utilized as a probe 
of kinematic substructure. This emission line is a prevailing spectral line in H {\scriptsize 
\textsc{II}} regions because its critical density ($6\times10^4$ cm$^{-3}$) is two orders 
of magnitude higher than the typical electron density of the galactic H {\scriptsize 
\textsc{II}} regions \citep{CMSC00}. Furthermore, photons emitted from the ions are 
not reabsorbed along the line of sight, and therefore this emission line is a useful tool 
to study the tomography of the ionized gas. Fig.~\ref{fig2} shows the comparison of 
[N {\scriptsize \textsc{II}}] $\lambda6584$ forbidden line with H$\alpha$ emission 
line for the same fibre position. 

The profile of the forbidden line [N {\scriptsize \textsc{II}}] $\lambda6584$ for 
some positions was found to contain a few emission components with different 
velocities. We obtained the RV of each component using a multiple Gaussian profile 
fitting method. A total of 1039 emission components were identified from the spectra 
of 480 fibre positions, and the RVs of 760 components were obtained with errors 
better than 1 km s$^{-1}$. On the other hand, the RVs of warm gas in the tadpole 
nebulae Sim 129 and 130 were measured from the 1 -- 0 S(1) emission line of hydrogen 
molecules at 2.12 $\mu$m. IGRINS provides about 60 near-infrared spectra along 
the position axis of the slit. We carried out single Gaussian profile fitting to the 
emission line for given positions and obtained the RVs of Sim 129 and 130 from 
342 and 469 positions with measurement errors better than 1 km s$^{-1}$, 
respectively. All the best fitting solutions were derived with the \textsc{MPFIT} 
packages \citep{M09}.

The spectra of late-type stars in the wavelength range of 5150 -- 5300 \AA \ contain 
a large number of metallic absorption lines as well as the prominent Mg b triplet 
at 5167, 5172, and 5183 \AA. The RVs of those stars can be precisely measured using 
a cross-correlation technique. We synthesized stellar spectra for the Solar abundance 
in the wide temperature range of 3800 $K$ to 9880 $K$ using the spectrum analysis 
code \textsc{MOOG} \citep{S73} and Kurucz ODFNEW model atmospheres \citep{CK04}. 
Synthetic spectra of B-type stars in the temperature range of 15000 $K$ to 25000 $K$ were 
obtained from Tlusty model atmospheres \citep{LH07}. These synthetic spectra were used 
as template spectra to derive the cross-correlation functions (CCFs) of the observed 
spectra. Since the spectral types of the observed late-type stars were 
unknown, we selected the synthetic spectrum with the strongest CCF peak among those 
of the other synthetic spectra as the template spectrum for an observed star. The velocities 
corresponding to a CCF peak were determined using \textsc{RVSAO} packages 
\citep{KM98}. The {\it xcsao} task provides the uncertainties of RVs based on the 
r-statistics as below \citep{TD79}:

\begin{equation}
r = {h \over \sqrt{2}\sigma_a}
\end{equation}

\noindent where $h$ and $\sigma_a$ are the amplitude of a CCF and the rms from 
its antisymmetric component, respectively. Then, the measurement error 
is expressed as $3w/8(1+r)$ where $w$ is the full width at half-maximum of the CCF 
peak \citep{KM98}. The mean error of RVs ($\sigma_{\mathrm{err}}$) in this study is 
about 1.3 km s$^{-1}$. We derived the CCFs of 119 stars, of which eight stars were identified 
as double-lined spectroscopic binary candidates. The spectra of 22 stars had insufficient 
signal-to-noise ratios to derive reliable CCFs, while, due to rapid rotation, another 42 A 
to B-type stars showed no lines or only a few broad, shallow absorption lines. 

O-type and early-B-type stars have a small number of weak metallic absorption 
lines in their spectra. Hydrogen lines in these stars are, in general, strong broad 
absorption lines that are often affected by the stellar wind. The cross-correlation 
technique is, therefore, less accurate for measuring the RV of O-type stars. However, 
a handful of helium lines between 4000 \AA \ and 5000 \AA \ can be used to determine 
their RV since the consistency of RVs from He {\scriptsize \textsc{II}} $\lambda\lambda$4200, 
4541, He {\scriptsize \textsc{I}} $\lambda\lambda$4387, 4713, and 4922 has been 
well tested \citep{Sdd13}. He {\scriptsize \textsc{I}} $\lambda\lambda$4387, 4713, and 4922 
are either too weak or absent in early-O-type stars, and virtually impossible to identify 
due to severe rotational broadening. Furthermore, these lines can be contaminated by the 
light from B-type companions. In the present work, He {\scriptsize \textsc{II}} 
$\lambda\lambda$4200 and 4541 were used to measure the RVs of four O-type stars, 
HD 242908, HD 242926, HD 242935, and TYC 2394-1214-1, while all five helium lines 
were used for BD +33 1025. The RV of the B0.2V star TYC 2394-1500-1 \citep{MJD95} was 
measured from He {\scriptsize \textsc{I}} $\lambda\lambda$4143, 4387, 4713, 
and 4921 as these lines have been used for rapidly rotating B-type stars \citep{EKD15}. The 
line centre of the helium lines for the six early-type stars was determined using a 
Gaussian profile fitting method and subsequently converted to velocity by comparing 
the wavelength of the line centre with the rest wavelength. We averaged the RVs 
measured at the same epoch.

\begin{figure*}
   \centering
   \includegraphics[width=15cm]{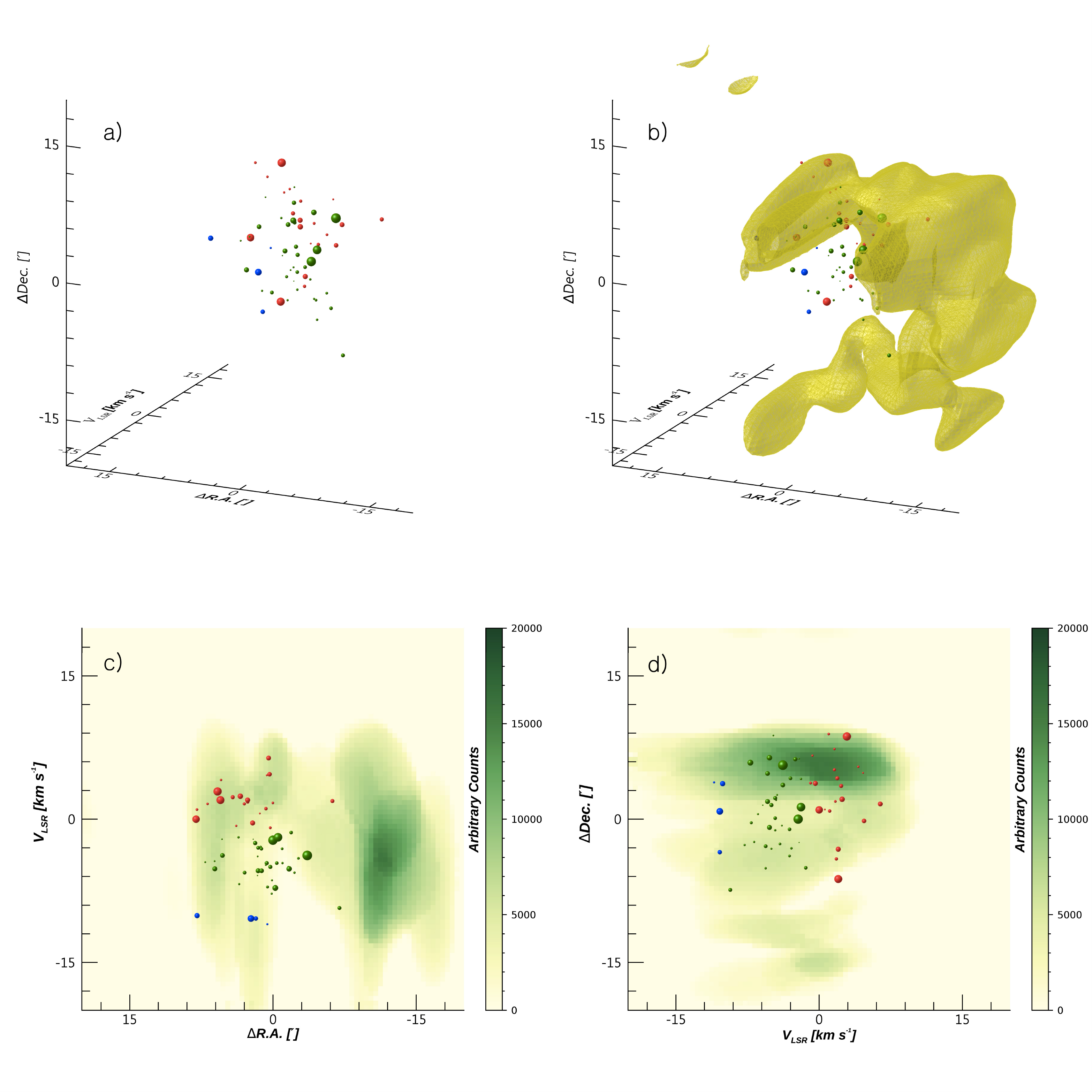}
\caption{Position-position-velocity diagrams ({\bf a} and {\bf b}\protect\footnotemark[2]) 
and position-velocity diagrams ({\bf c} and {\bf d}) of NGC 1893. Isosurface and colour 
contours represent the distribution of ionized gas traced by the forbidden line [N {\scriptsize 
\textsc{II}}] $\lambda6584$, where the unit of the counts is electrons ($e^{-}$). Spheres 
denote the stars in the cluster, and their size is proportional to the brightness of individual 
stars. Blue, green, and red colours represent three different groups in the radial velocity range 
of $V_{\mathrm{LSR}} = -19$ km s$^{-1}$ to $-10$ km s$^{-1}$, $-10$ km s$^{-1}$ 
to $-1$ km s$^{-1}$, and $-1$ km s$^{-1}$ to 8 km s$^{-1}$, respectively. The 
position of objects is relative to the O-type star HD 242935 (R.A. $= 05^{\mathrm{h}} 
\ 22^{\mathrm{m}} \ 46.^{\mathrm{s}}5$, Dec. $= +33^{\circ} \ 25^{\prime} \ 
11^{\prime\prime}$, J2000).}
         \label{fig3}
   \end{figure*}
\footnotetext[2]{The animation of this figure can be accessed through the links \url{http://starburst.sejong.ac.kr/data/fig-3b_ani.gif} and \url{http://starburst.sejong.ac.kr/data/N1893_ppv_ul.gif}}

Since the O-type stars BD +33 1025, HD 242908, HD 242926, and HD 242935 
were observed in two different epochs, the variability of RVs could be checked roughly. 
BD +33 1025 showed a variation of 12.2 km s$^{-1}$ for 47 days, and this star could 
be a single-lined spectroscopic binary. The RV of HD 242908 was almost constant for 
48 days. This star could be either a single star or long period binary system. A variation 
of 4.2 km s$^{-1}$ was found for HD 242935 in the centre of the cluster. Since these 
variations are comparable to the typical uncertainty of RV for O-type stars (a few km s$^{-1}$), 
further observations are required to identify binaries among these O-type stars. On the 
other hand, we found a significant variation of line profiles for He {\scriptsize \textsc{I}} 
$\lambda\lambda$4387, 4713, and 4922 in the spectra of HD 242926. This variation 
is most likely caused by light from a B-type companion star. Indeed, the RV of this star 
significantly changed by up to 61.8 km s$^{-1}$ over 39 days. Hence, HD 242926 is a certain 
double-lined spectroscopic binary. All the double-lined spectroscopic binary candidates 
identified in this study were not used for further analysis. The RV measurements from 
this work are available in the electronic tables (Table~\ref{tab2} and \ref{tab3}) or from 
the author (BL).

\begin{figure*}
   \centering
   \includegraphics[width=15cm]{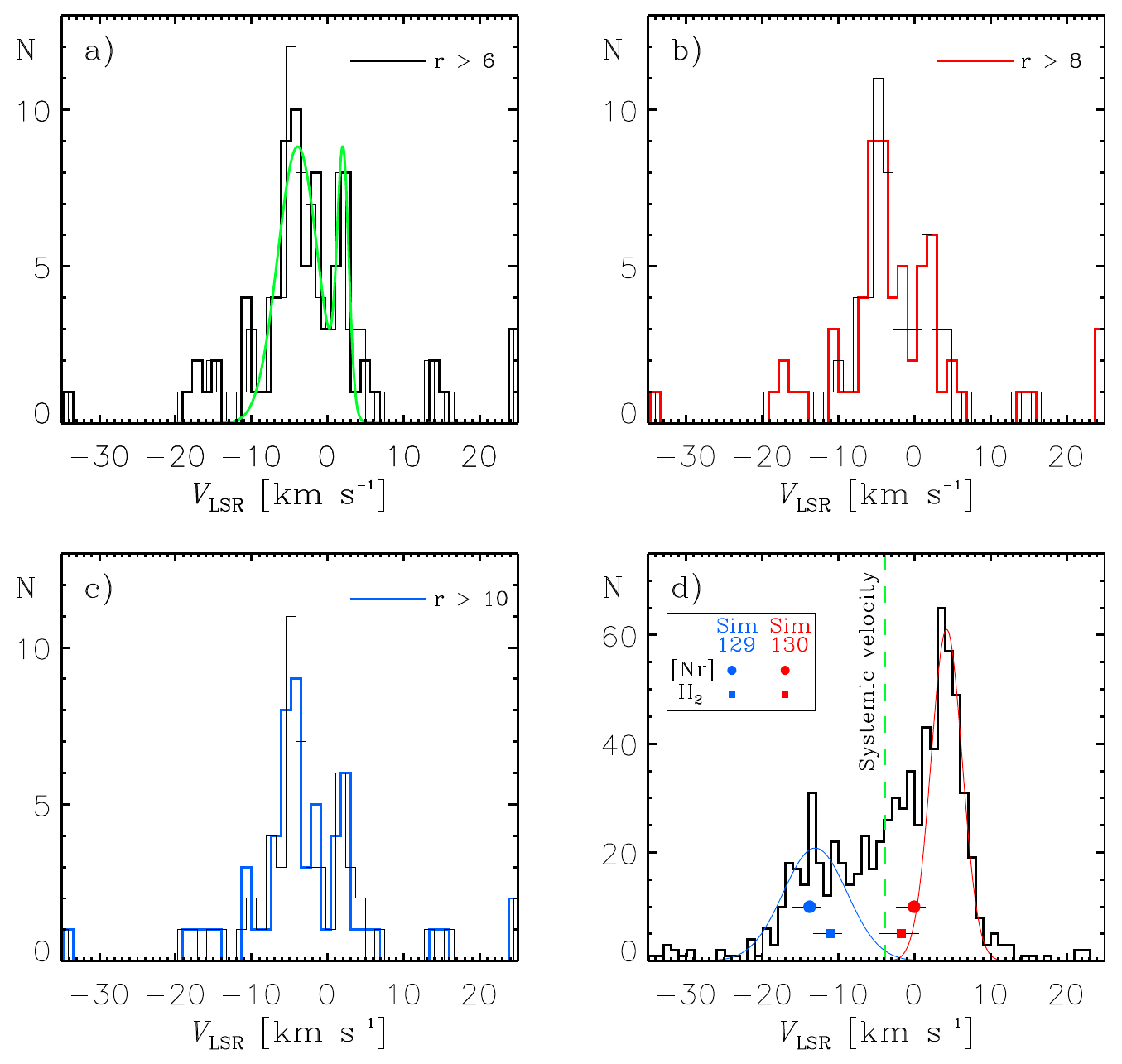}
      \caption{Radial velocity distributions of stars ({\bf a -- c}) and gas ({\bf d}). Black, 
red, and blue histograms represent the radial velocity distributions for subsets of stars 
with r-statistics larger than 6, 8, and 10, respectively. The bin size is 1.3 
km s$^{-1}$, equivalent to the mean measurement error. To avoid the binning effect, 
we also overplotted histograms (thin solid line) shifted by 0.65 km s$^{-1}$. The 
radial velocity distribution of stars was fitted by a double-Gaussian profile. The 
histogram ({\bf d}) exhibits the radial velocities of ionized gas traced by the forbidden 
line [N {\scriptsize \textsc{II}}] $\lambda6584$. The dashed line represents the systemic 
velocity of the cluster. The mean velocities of each bubble obtained from a Gaussian 
profile fitting method are about $-13.0 \pm 1.4$ and $4.2 \pm 0.1$ km s$^{-1}$, 
respectively. The circles show the mean velocities of the ionized component in the tadpole 
nebulae Sim 129 (blue) and 130 (red), while the squares denote those of their warm components 
measured from the 1 -- 0 S(1) emission line of hydrogen molecules. The error bars indicate the 
minimum and maximum values of their radial velocities.}
         \label{fig4}
   \end{figure*}

\section{Kinematic substructure}
The colour-composite images of NGC 1893 show the spatial distribution of hot ionized 
gas and stars (Fig.~\ref{fig1}). A total of five O-type stars are distributed 
in a north-south direction, two of which are located in the cluster centre. The intracluster 
medium is filled with hot ionized gas as seen in the [O {\scriptsize \textsc{III}}] 
$\lambda5007$ image. The tadpole nebulae Sim 129 and 130 facing towards the 
two O-type stars in the cluster centre are glowing in the narrow band H${\alpha}$, 
H${\beta}$, and [S {\scriptsize \textsc{II}}] $\lambda6712$ filter images as shock heated 
gas-flows surround these nebulae. 

We present two and three-dimensional views of this cluster in position-velocity 
space as shown in Fig.~\ref{fig3}. In order to display the distribution of the ionized gas 
in the diagrams, a data cube made of $80\times80\times80$ volume cells was created by 
linearly interpolating the scattered data values to points in a regularly sampled volume. 
In this procedure, the observed parameters, such as the fibre positions (right ascension and 
declination), the velocity profiles of the [N {\scriptsize \textsc{II}}] emission line, and the 
counts ($e^{-}$) which were irregularly scattered in a three-dimensional space were 
interpolated to a regular grid cube by applying Delauney triangulation technique. The 
two-dimensional position-velocity diagrams were obtained by the sum of the counts 
along one position axis (either right ascension or declination) in the data cube. 

The isosurface shows the distribution of the ionized gas that surrounds the whole 
cluster like a bubble (Fig.~\ref{fig3}-b). Such a gas bubble appears open towards 
observers in the position-velocity diagrams (c and d). Its asymmetric shape is attributed 
to the inhomogeneous distribution of matter. The fact that the stars and ionized gas 
coexist in position-velocity space, indicates that they are physically associated 
components in NGC 1893. The ionizing source is O-type stars rather than either 
supernova explosions of the most massive stars, or violent activities arising from 
compact objects, because NGC 1893 is too young to produce the sources that could 
cause such energetic events.

\begin{figure*}
   \centering
   \includegraphics[height=0.3\textwidth]{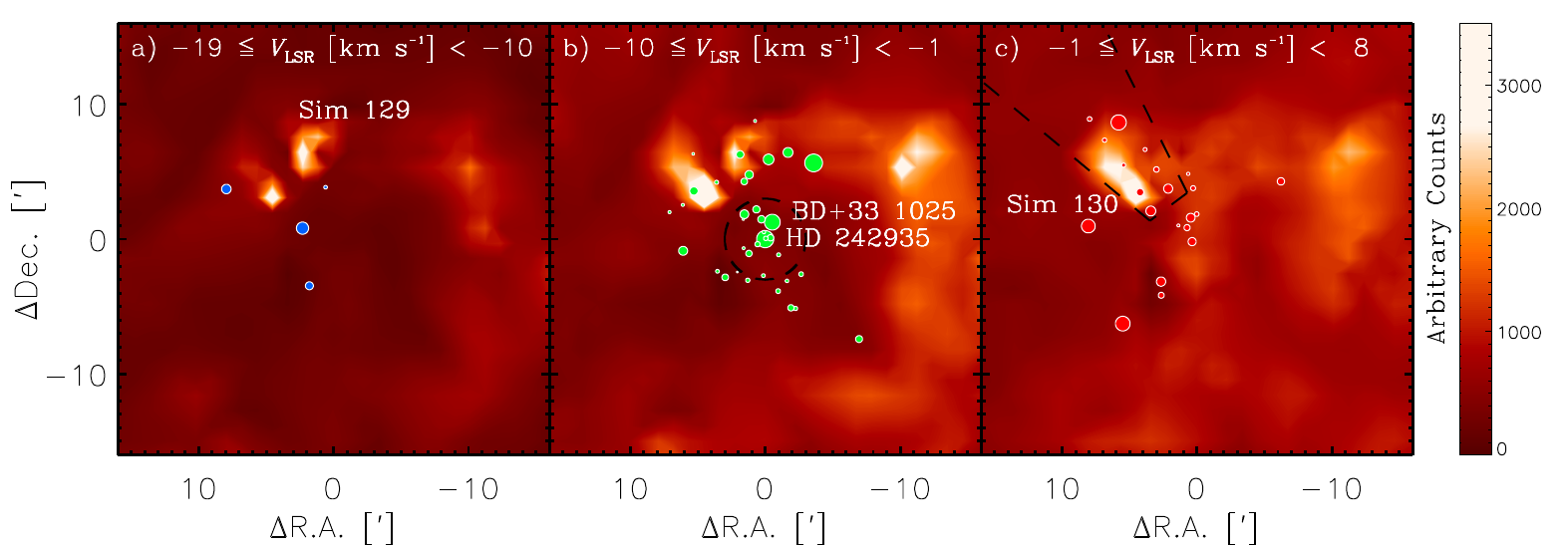}
      \caption{{\bf a -- c}, integrated channel maps of ionized gas across NGC 1893. These maps 
were obtained by integrating all the counts of the forbidden line [N {\scriptsize \textsc{II}}] 
$\lambda6584$ within the given velocity ranges. Stars in the same velocity ranges as the 
ionized gas are also plotted. The size of circles is proportional to the brightness of 
individual stars. The coordinates relative to the cluster centre are the same as those 
in Fig.~\ref{fig3}. The circle (dashed line) in {\bf b} represents the size of the central 
cluster inferred from the radial surface density profile of stars (3$^{\prime}$), and 
the kinematic subgroup is confined by a region outlined by dashed lines in {\bf c}. }
         \label{fig5}
   \end{figure*}

The RVs of the stars show a bimodal distribution (Fig.~\ref{fig4}-{\bf a}). The possibility 
that the observed distribution was from a single Gaussian distribution is less than 
10 per cent according to a Kolmogorov-Smirnov test. The bimodality is 
conserved even when the r-statistics is changed or when the histograms were shifted 
by the half size of the adopted velocity bin. We also investigated the histograms by 
changing the bin sizes of $n\sigma_{\mathrm{err}}$ (where $n = 0.5, 1.0, 1.5, 2.0,$ 
and 2.5). The components of the subgroup were clearly seen in the histograms
(with confidence levels at least two times higher than the Poisson noise of the histogram 
of the central cluster) until with the bin size of 2.5$\sigma_{\mathrm{err}}$ they 
merged into that of the central cluster. Hence it is certain that there are, at least, 
two groups of stars with different velocity fields in the line of sight. 

We fitted a double-Gaussian profile to the observed RV distribution 
as seen in Fig.~\ref{fig4}-{\bf a}. The main peak in the histogram indicates 
the systemic velocity of NGC 1893. The systemic velocity of the cluster was estimated 
to be $-3.9 \pm 0.4$ km s$^{-1}$. The dispersion of the systemic velocity 
is about 2.6 km s$^{-1}$. Given a mean measurement error of 1.3 km s$^{-1}$, 
as well as the contribution of binaries with small velocity variations, the intrinsic 
velocity dispersion is likely to be smaller than 2.3 km s$^{-1}$. The 
RV of the subgroup corresponding to the secondary peak is about $+2.0
\pm0.3$ km s$^{-1}$, and its dispersion is about 0.8 km s$^{-1}$ which is smaller 
than the measurement error. The small dispersion may be attributed to the small 
number of sample stars.

The RV distribution of the ionized gas also exhibits a non-Gaussian distribution 
(Fig.~\ref{fig4}-{\bf d}), as can be expected from its bubble structure. The well-defined systemic 
velocity from the RV distribution of stars almost equally divides the gas bubble into a near 
side part and a far side part. The component with a low-amplitude at $-13.0$ km s$^{-1}$ 
is indicative of a small amount of material on the near side of the bubble. On the other hand, 
the component with a high-amplitude at 4.2 km s$^{-1}$, corresponds to material on the 
far side of the bubble where ionization fronts proceed into the remaining molecular cloud. 
The relative RV of each peak compared to the systemic velocity indicates the ionized bubble 
has an expansion velocity of 8.1 -- 9.1 km s$^{-1}$ from the cluster centre. The mean RVs 
of the ionized gas and warm hydrogen molecules in the tadpole nebulae show that they lie 
at different parts of the ionized gas bubble. 

Sim 129 is a clump in the direction of the near side of the bubble (Fig.~\ref{fig4}-{\bf d}). No star 
formation has taken place in Sim 129 given that no star is kinematically associated with the 
nebula (Fig.~\ref{fig5}-a). The largest rim of ionized gas seen in the integrated channel map 
Fig.~\ref{fig5}-b is the section of a bubble. If the gas bubble has been being expanding 
at a mean velocity of 8.6 km s$^{-1}$ for 1.5 Myr at 3.5 kiloparsecs, the size of the 
rim should be about 26$^{\prime}$ in diameter. This estimate is in reasonable agreement 
with the size of the observed bubble. Stars in the same velocity range are distributed 
across the whole region. Since HD 242935 and BD +33 1025 in the cluster centre have velocities 
within the systemic velocity range, they are major contributors to the creation of the central 
cavity and the tadpole nebulae. The most prominent feature in Fig.~\ref{fig5}-{\bf c} is the distribution 
of stars correlated with the head-tail structure of Sim 130 (see the region outlined 
by dashed lines). This kinematic substructure is unlikely to result from the dynamical 
evolution because the age of the cluster is 2.5 times younger than the crossing time of 
the stars (${8.9 \ \mathrm{pc} \over 2.3 \ \mathrm{km \ s^{-1}}} \sim $3.8 Myr, see 
Section 6). We find, therefore, that this group is spatially and kinematically associated with Sim 130.  

\begin{figure}
   \centering
   \includegraphics[width=7cm]{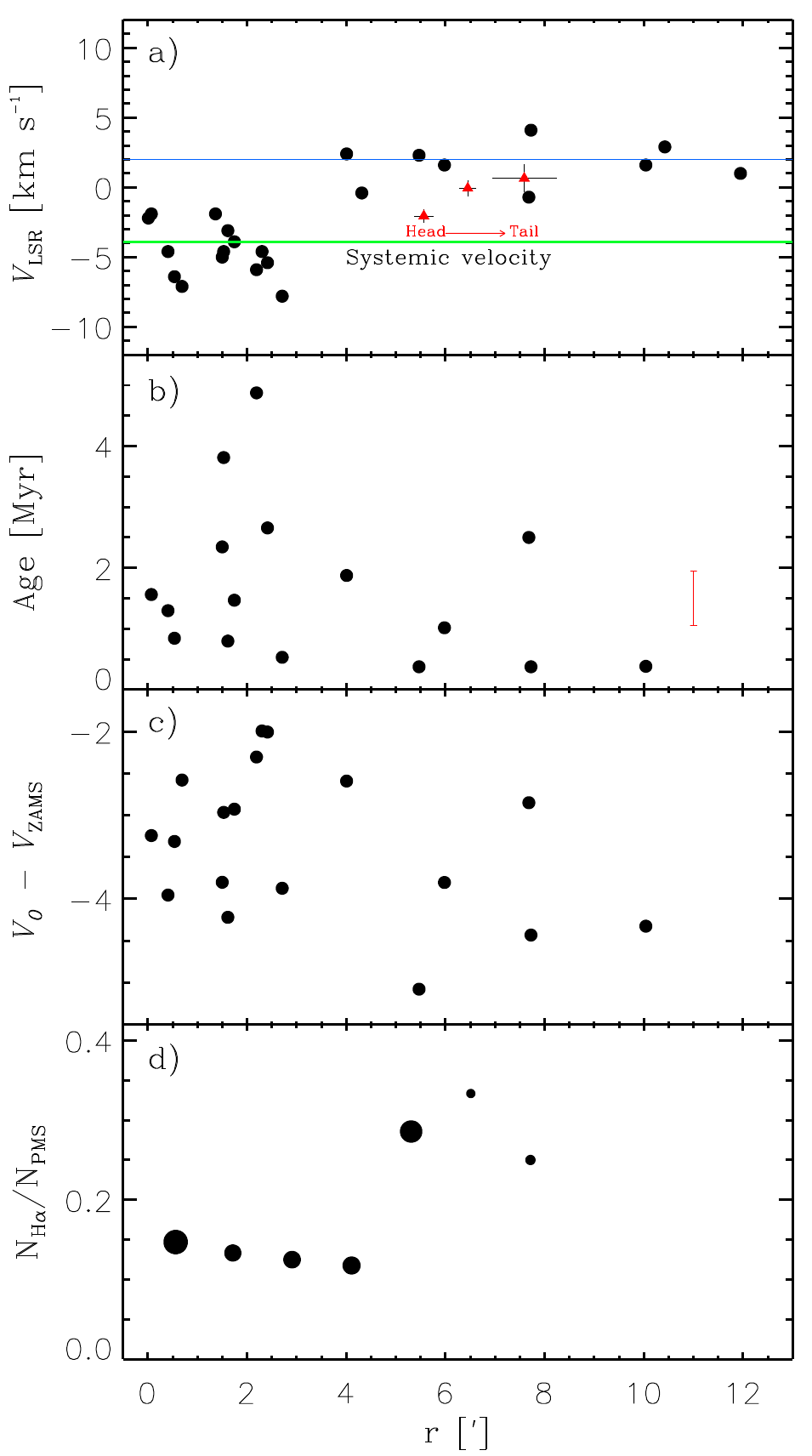}
      \caption{Radial variation of the properties of pre-main sequence 
stars in the regions outlined by dashed lines in Fig.~\ref{fig5}-{\bf b} and {\bf c}. {\bf (a)} The radial 
velocity distribution of the stars. Blue thin and green thick solid lines represent the mean 
radial velocity of newborn stars and the systemic velocity of the cluster, respectively. 
Red triangles shows radial velocity variation along the head-tail structure of Sim 130. 
{\bf (b)} The age distribution of the stars inferred from the evolutionary models for 
pre-main sequence stars \citep{SDF00}. The error bar denotes the size of the typical 
error. {\bf (c)} The brightness distribution of pre-main sequence stars. $V_0 - 
V_{\mathrm{ZAMS}}$ represents the difference of visual magnitude between the pre-main 
sequence stars and the zero-age main-sequence relation at given colours. {\bf (d)} The 
variation of the fraction of H$\alpha$ emission stars. The size of the circles are proportional 
to the number of pre-main sequence stars. See the main text for details.}
         \label{fig6}
   \end{figure}

A velocity variation of 2.7 km s$^{-1}$ was found along the head-tail structure of Sim 130 
(see Fig.~\ref{fig6}-{\bf a}). The ionized gas tail has almost the same velocity field as that of the 
subgroup, whereas its head is moving away more slowly than its tail from the cluster centre. 
Given that neither remarkable acceleration nor deceleration was found in the RVs of the stars 
in the subgroup, the velocity variation probably does not represent the acceleration history 
of the ionized gas bubble. Irradiation on an inhomogeneous molecular cloud is likely 
responsible for the formation and evolution of Sim 130 \citep{GNW09}.

\section{Age trend}
The subgroup in the vicinity of Sim 130 is expected to be younger than 
the central cluster if star formation was triggered by feedback. In order to confirm 
the age difference between the central cluster and the subgroup, the age distribution 
of PMS stars was investigated using our large photometric sample. Their age was inferred 
from the Hertzsprung-Russell diagram by comparing their effective temperature and 
bolometric magnitude with those of evolutionary models (see \citealt{LSK14} for detail). 
PMS stars with a photometric error smaller than 0.1 mag were used to obtain a reliable 
age distribution. The errors in the visual magnitude, $V-I$ colour, reddening correction, 
distance modulus, and bolometric correction were propagated. Apart from the errors in 
observations and calibrations, the duration of a star-forming event can cause an additional 
broadening of the age distribution. While there could be a complicated star-formation 
history in the whole region, the star-formation timescale in a small region, such as the 
central cluster, may be negligible compared to that of the whole region. We, therefore, 
restricted our sample to PMS stars within the radius of the central cluster (3$^{\prime}$) 
inferred from the radial surface density profile of stars. 

The derived age distribution showed a broad asymmetric Gaussian profile with a strong 
peak at 1.5 Myr corresponding to the age of the central cluster. Given that the 
age distribution is almost the same as the distribution of the propagated error, the 
dispersion of the age distribution represents the error in age estimation. We 
conducted a skew Gaussian profile fitting to this age distribution. The typical error in 
age was estimated to be 0.9 Myr from the dispersion ($\sqrt{\mathrm{variance}}$) 
of the best-fit profile. However, the age error of individual stars could be larger than 
the typical error due to the imperfect correction of differential reddening and systematic 
uncertainties in evolutionary models for 1.5 -- 2 solar-mass PMS stars \citep{SBC04}.

We plotted the age of the PMS stars within two regions outlined by 
dashed lines in Fig.~\ref{fig5}-{\bf b} and {\bf c} with respect to the distance 
from the cluster centre (Fig. ~\ref{fig6}-{\bf b}). Since the age distribution 
has a long tail from 5 Myr towards old age, PMS stars younger than that age were 
used to seek a clear age gradient. Interestingly, in the subgroup of stars in the 
direction of Sim 130, the age of the PMS stars tend to be younger than that of stars 
in the cluster centre. This age difference between stars in the vicinity of Sim 130 
and the cluster centre has been reported in other work \citep{SPO07,MSB07,PSC13} 
as well as in our previous work \citep{LSK14}. However, since the age difference is 
comparable to the typical error, this result does not guarantee the robustness of the age 
difference between them.

We tried to test two different age indicators. The luminosity of 
PMS stars decreases as they contract along their Hayashi tracks, and therefore 
younger PMS stars are brighter than older stars at a given temperature. The extinction 
of individual PMS stars in visual magnitude $V$ was corrected using the reddening 
map \citep{LSK14}, and then the extinction-corrected visual magnitude ($V_0$) was 
compared with that ($V_{\mathrm{ZAMS}}$) of the zero-age main sequence relation at 
given colours \citep{SLB13}. Fig.~\ref{fig6}-{\bf c} shows the radial variation of 
$V_0 - V_{\mathrm{ZAMS}}$. The lower the value, the younger the star. Star ID 
6091 (R.A. = $05^{\mathrm{h}} \ 23^{\mathrm{m}} \ 24.^{\mathrm{s}}72$, 
Dec. = $+33^{\circ} \ 34^{\prime} \ 05.^{\prime\prime}8$) was found near 
the main sequence band although it was identified as a Class II object \citep{CMP08}. 
This star may have a nearly edge-on disc as did those 
found in the young open cluster NGC 2264 \citep{SSB09}. We excluded this star 
in this analysis. The median values of $V_0 - V_{\mathrm{ZAMS}}$ for the central 
cluster and the subgroup are $-3.0$ and $-3.8$, respectively. The subgroup seems 
to be slightly younger than the central cluster.

Since the number of strong H$\alpha$ emission stars which can be identified 
by H$\alpha$ photometry rapidly declines as a function of time, the fraction 
of H$\alpha$ emission stars can be used as an age indicator. Seven 
circular regions with the same area (0.79 square arcminutes) were chosen along 
the line joining the central cluster and Sim 130. The number of H$\alpha$ emission 
stars and that of all the identified PMS stars were counted within given areas, 
respectively, and then the fraction of H$\alpha$ emission stars were computed 
for each area. Fig.~\ref{fig6}-{\bf d} shows that the fraction in the vicinity of Sim 
130 is about two times higher than that in the central cluster. Although 
H$\alpha$ emission can vary with time depending on the degree of accretion 
activity, such a distinctive difference between the two regions cannot be explained 
by variability. These results are indicative of an age difference and support 
the fact that the subgroup associated with Sim 130 is a new generation of stars 
formed in the expanding gas due to feedback from the O-type stars. 

\begin{figure*}
   \centering
  \includegraphics[width=5.3cm]{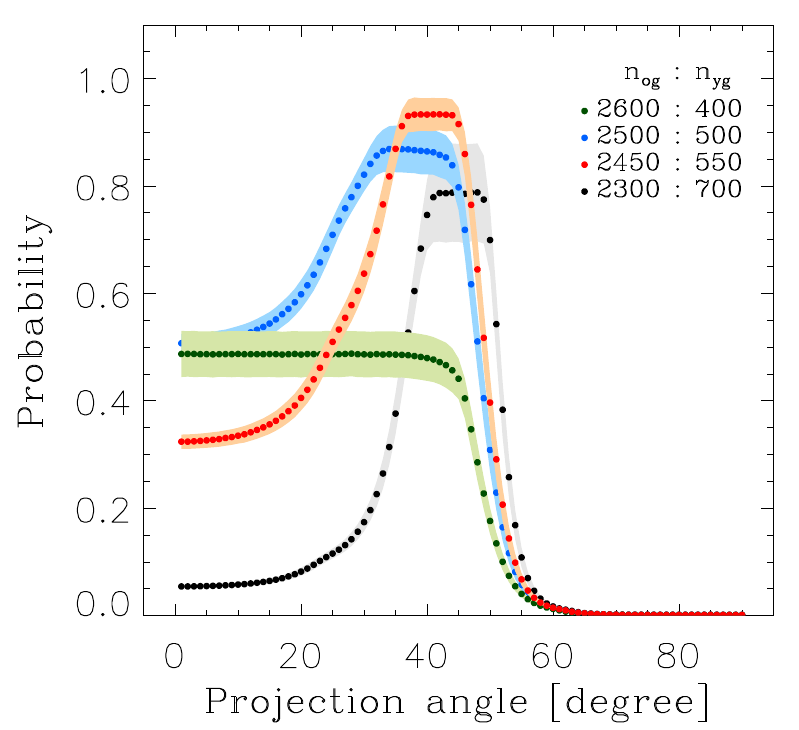}  \includegraphics[width=5.3cm]{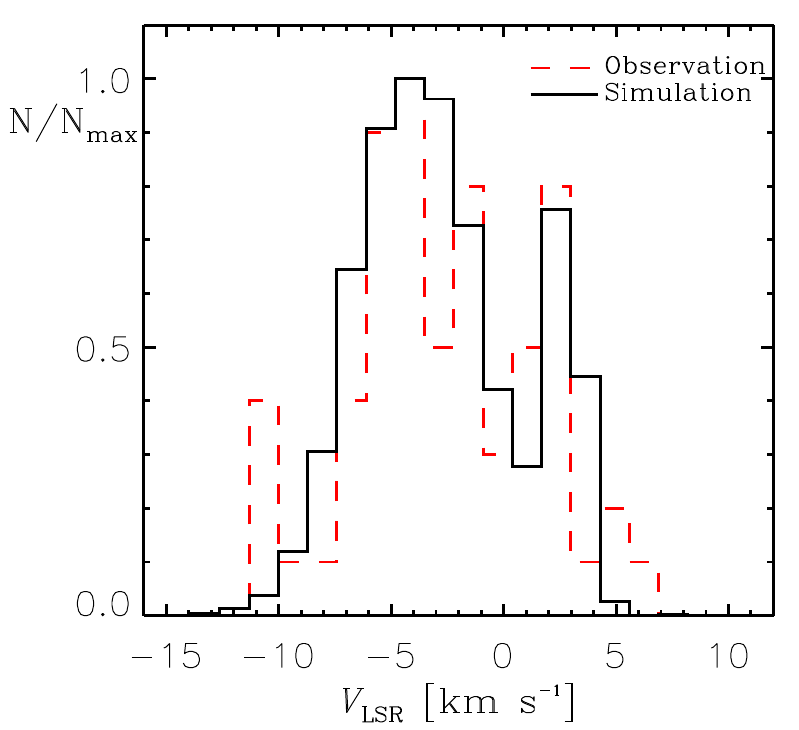}\includegraphics[width=5.3cm]{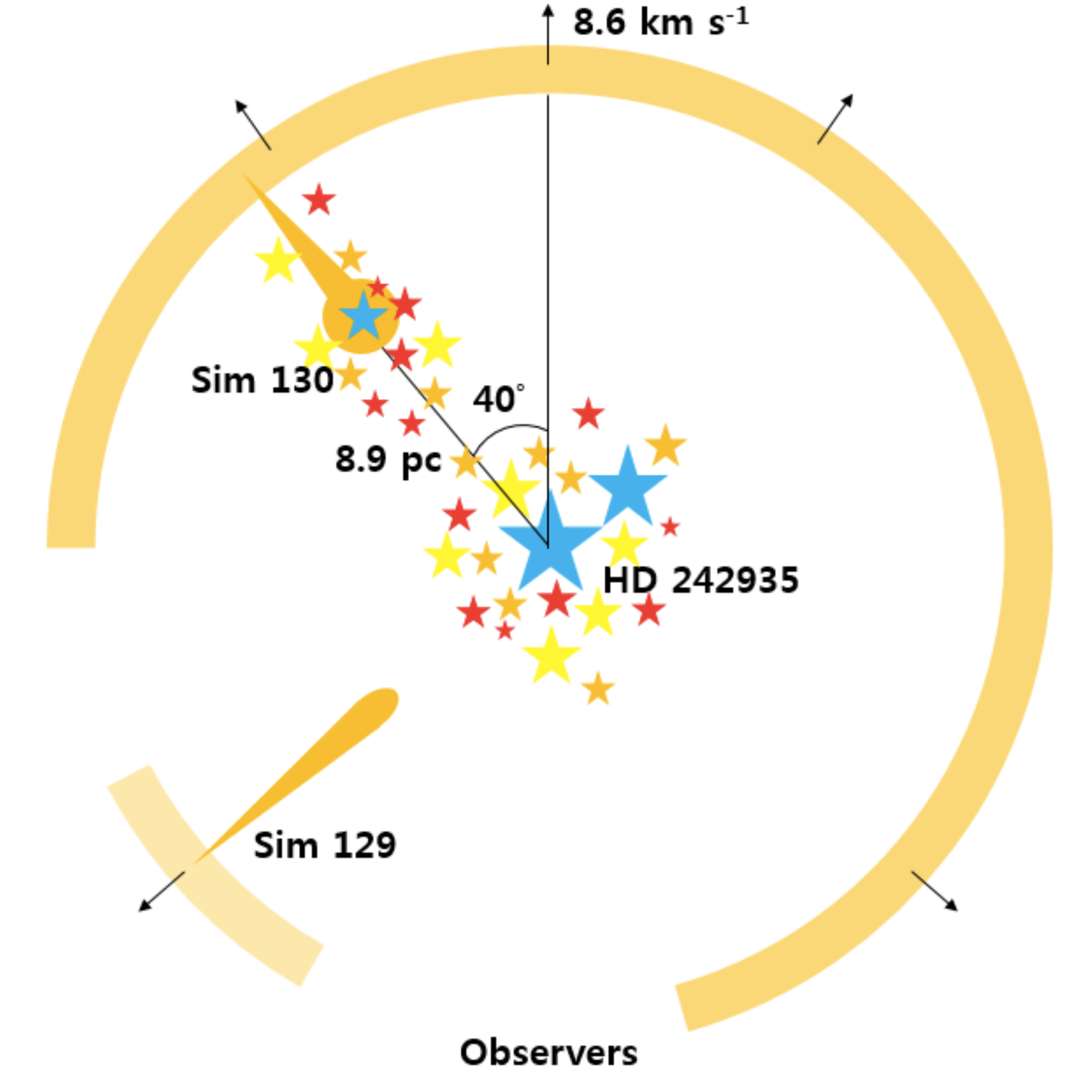}
      \caption{Probability functions of projection angles to the line 
of sight (left), comparison of the best model with the observed distribution (middle), 
and schematic figure overlooking NGC 1893 (right). The projection angle is defined 
as the angle between the kinematic subgroup-HD 242935-the line of sight. The mean 
probabilities were plotted by dots, and their 1$\sigma$ errors were shown by shaded 
regions. Different number ratios between old and young groups were applied for the 
simulations as shown by different colours. The best solution with the highest confidence 
level is the model cluster adopting a number ratio of 2450:550 ($n_{\mathrm{og}}$:$n_{\mathrm{yg}}$) 
at a projection angle of about $40\pm2^{\circ}$ (s. d.). The schematic figure exhibits the internal structure 
of the cluster reconstituted from stellar and gas kinematics. A huge ionized gas bubble 
is expanding at 8.1 -- 9.1 km s$^{-1}$ (on average 8.6 km s$^{-1}$) away 
from the central O-type stars. Dense parts of a molecular cloud can remain as gas 
pillars as the massive stars blow out the remaining gas. Depending on the local gas 
density distribution, some of the dense pillars compressed by ionization and shock 
fronts will form a new generation of stars.}
         \label{fig7}
   \end{figure*}

The Taurus-Auriga T association covers a 40 $\times$ 40 square parsec area and is an ideal 
comparison field because, low-mass star formation is dominant in the absence of massive O-type 
stars. The subgroups of young stars show substructure which is considered to be inherited 
from the filaments of their natal clouds \citep{ADW14} and share almost the same kinematic 
properties with each other \citep{BG06}. The stars of the subgroups do not 
show any systematic difference \citep{PS02} within 4 Myr, and therefore the 
formation of the subgroups in the Tau-Aur T association seems to have occurred at 
almost the same time. The kinematic and photometric properties of stars in NGC 1893 
are far different from those of stars in the Taurus-Auriga T association. Hence, the newborn 
stars in the subgroup with a receding velocity from the central cluster are not stars spontaneously 
formed at a recent epoch.

\section{Contribution of the new generation of stars}
NGC 1893 is composed of about 3000 stars according to the intrinsic number of the 
cluster members inferred from its initial mass function \citep{LSK14}, and the members 
are divided into at least two groups with different velocity fields; an old group with the 
systemic velocity of NGC 1893 and a young group. The young group is moving away 
from the old group at a specific projection angle, where the projection angle is defined 
as the angle between the young group-HD 242935-the line of sight (see Fig.~\ref{fig7}). 
The stars in the young group will then have RVs as below:

\begin{equation}\label{eq:2}
V_{\mathrm{obs}} = V_{\mathrm{sys}} + V_{\mathrm{exp}} \times \mathrm{cos}\theta_{\mathrm{pro}}
\end{equation}

\noindent where $V_{\mathrm{sys}}$, $V_{\mathrm{exp}}$, and $\theta_{\mathrm{pro}}$ 
are the systemic velocity of NGC 1893, the expanding velocity of the ionized gas bubble, 
and a projection angle, respectively. Here, the ionized gas bubble was assumed to be isotropically 
expanding from the central cluster at 8.6 km s$^{-1}$ which is the mean expanding velocity of 
the H {\scriptsize \textsc{II}} bubble. Hence, the observed RV distribution of stars depends 
on the geometric distribution of stars within NGC 1893. In addition, the relative amplitudes 
of the observed RV distribution between the old and young groups are related to their number 
ratio. We performed multiple sets of simulations to obtain the geometry of NGC 1893 and the 
number ratio.

Artificial stars in the old group were set to have RVs drawn from a Gaussian distribution 
with the systemic velocity of $-3.9$ km s$^{-1}$ and the observed velocity dispersion of 
2.6 km s$^{-1}$ using a Monte-Carlo method. The RVs of artificial stars in the young group 
were obtained by substituting various projection angles from 0$^{\circ}$ to 90$^{\circ}$ 
with a 1$^{\circ}$ interval to the Eq.~\ref{eq:2}, and the velocity dispersion of 0.8 km s$^{-1}$ 
was applied to the RV distributions of this group. Arbitrary number ratios between the 
old and young groups were assumed to be $n_{\mathrm{og}}$:$n_{\mathrm{yg}}$ = 2300:700, 
2400:600, 2500:500, 2600:400 and 2700:300. A given run for each projection angle and number 
ratio was repeated 10,000 times, and the resultant distributions were compared with 
the observed RV distribution, where stars satisfying two criteria; 1) r-statistics $> 6$, 
2) $-12 < V_{\mathrm{LSR}} \ [\mathrm{km s^{-1}}] < 12$, were used. The probabilities, 
that represent similarity between the RV distributions from observation and simulations, were 
computed using a Kolmogorov-Smirnov test, and these were averaged for the 
same parameter setup. The highest mean probability among these runs was found 
between $n_{\mathrm{og}}$:$n_{\mathrm{yg}} $= 2400:600 and 2500:500. We 
made fine adjustments by changing the number ratios to $n_{\mathrm{og}}$:$n_{\mathrm{yg}} $
= 2440:560, 2450:550, and 2460:540. The model cluster adopting $n_{\mathrm{og}}$:$n_{\mathrm{yg}}$ = 2450:550 well reproduced the observed RV distribution at $40\pm 2^{\circ}$ (s. d.) 
with a 93 per cent confidence level (Fig.~\ref{fig7}). 
 
We obtained the proper distance of 8.9$^{+0.39}_{-0.35}$ parsecs between the head of Sim 130 and the 
O-type star HD 242935 from the projected distance (5.7 parsecs) and the projection 
angle using a trigonometric function. The timescale of feedback-driven star formation 
is about 1 Myr assuming a mean expanding velocity of 8.6 km s$^{-1}$. 
The triggered population accounts for about 18 per cent of all the cluster members according 
to the number ratio used in the best model. This contribution seems to be a lower limit 
because we did not take into account the motion perpendicular to the line of sight. 
Our results suggest that feedback from massive stars in a core cluster can play an important 
role in building-up the stellar content of OB associations.

\section{Summary and conclusion}
In the present work, we tested the self-regulating star formation model proposed 
by \citet{EL77} using high resolution spectroscopy of stars and gas in the 
young open cluster NGC 1893. The results are summarized as below:

   \begin{enumerate}
      \item Narrow band images show that hot gas fills the intracluster medium 
as seen in the [O {\scriptsize \textsc{III}}] $\lambda5007$ filter image, and that shock-heated 
gas surrounds the tadpole nebulae Sim 129 and 130.

      \item The three-dimensional position-position-velocity and position-velocity diagrams 
show that an ionized gas bubble surrounds the cluster members, and that the gas and 
stars are globally associated with each other. The systemic velocity of NGC 1893 is about 
$-3.9$ km s$^{-1}$ and the ionized gas bubble is expanding from the cluster centre at 
8.1 -- 9.1 km s$^{-1}$.

      \item Sim 129 and 130 were found to lie on different parts of the bubble moving away 
from the cluster centre. There is no sign of star formation in Sim 129, while a subset of 
stars are coincident with Sim 130 in position-velocity space. Furthermore, as the stars in the 
vicinity of Sim 130 tend to be younger than those in the cluster centre, the stars associated 
with Sim 130 are likely a new generation of stars formed by feedback from massive stars.

       \item The geometry of NGC 1893 and the number ratios between old and young groups 
were inferred from multiple sets of Monte-Carlo simulations. The model adopting the 
projection angle of 40$^{\circ}$ and $n_{\mathrm{og}}$:$n_{\mathrm{yg}}$ = 2450:550 best
reproduced the observed RV distribution with a 93 per cent confidence level. The timescale of 
sequential star formation within the proper distance of 8.9 parsecs was estimated to be 1 
Myr. The contribution of the newborn stars to the total stellar population in 
NGC 1893 turned out to be at least 18 per cent.
   \end{enumerate}

These results indicate that feedback from massive stars can play an important 
role in the formation of OB association and supports the self-regulating star formation model \citep{EL77}. 
The forthcoming Large Synoptic Survey Telescope will observe the entire regions of a number of 
OB associations providing deep photometric data for their members, while the {\it Gaia} mission 
will provide the precise proper motions and distances for many individual stars. Data from such extensive 
surveys, combined with high resolution spectroscopy, will make unprecedented synergy in uncovering 
the formation process of OB associations in the near future.


\section*{Acknowledgements}
The authors thank the anonymous referee for many useful comments and also 
thank Perry Berlind, Mike Calkins, and Nelson Caldwell (Smithsonian Astrophysical 
Observatory) for assisting with Hectochelle observations, Wonseok Kang, Taewoo Kim, 
and Shinu Jeong (Deokheung Optical Astronomy Observatory) for narrow band imaging observations. 
This paper has used the data obtained under the K-GMT Science Program (PID: 16A-MMT-001) 
funded through Korean GMT Project operated by Korea Astronomy and Space Science Institute 
(KASI). In addition, the data taken from the Harlan J. Smith Telescope at the McDonald Observatory 
of The University of Texas at Austin (UT), as well as the Discovery Channel Telescope (DCT) 
at Lowell Observatory were used in this paper. Lowell is a private, non-profit institution dedicated to 
astrophysical research and public appreciation of astronomy and operates the DCT in partnership 
with Boston University, the University of Maryland, the University of Toledo, Northern Arizona 
University and Yale University. This work used the IGRINS that was developed under a collaboration 
between the UT and the KASI with the financial support of the US National Science Foundation 
under grant AST-1229522, of the UT, and of the Korean GMT Project of KASI. The resources 
of SELab were also used in this work. This work was supported by KASI grants 2017186000 
and 2017183003. B.L. and H.S. acknowledge the support of the National Research Foundation 
of Korea, Grant No. NRF-2017R1A6A3A03006413 and NRF-2015R1D1A1A01058444, respectively.






\begin{table*}
\begin{minipage}{150mm}
\caption{Radial velocities of gas measured from H$_2$ 1-0 S(1) emission line and the forbidden line [N {\scriptsize \textsc{II}}] $\lambda6584$.}
\begin{tabular}{ccccl}
\hline
R.A. (J2000) & Dec. (J2000)& $V_{\mathrm{LSR}}$ & $\epsilon V_{\mathrm{LSR}}$ & Remark \\
(h:m:s) & ($^{\circ}:^{\prime}:^{\prime\prime}$) & (km s$^{-1}$) & (km s$^{-1}$) &  \\
\hline
{\it IGRINS} &&&& \\
\hline
05:22:54.36 & $+$33:31:52.9 &$ -10.4$ & 0.4 & Sim 129 \\
05:22:54.37 & $+$33:31:52.9 &$-10.8$ & 0.3 & Sim 129 \\
05:22:54.39 & $+$33:31:52.9 &$ -11.6$ & 0.2 & Sim 129 \\
05:23:07.60 & $+$33:29:03.2 &$  -2.4$ & 0.3 & Sim 130 \\
05:23:07.61 & $+$33:29:03.2 &$  -2.5$ & 0.3 & Sim 130 \\
05:23:07.63 & $+$33:29:03.2 &$  -2.2$ & 0.3 & Sim 130 \\
\hline
{\it Hectochelle} &&&&\\
\hline
05:21:08.82 & $+$33:22:58.6 &$  -0.7$ & 0.7 & \\
                    &                          &$   6.9$ & 0.4 & \\
05:21:48.41 & $+$33:21:08.9 &$  -6.6$ & 0.4 & \\
\hline
\label{tab2}
\end{tabular}
\end{minipage}
\begin{tabular}{@{}l@{}}
The full table is available electronically.
\end{tabular}
\end{table*}

\begin{table*}
\begin{minipage}{150mm}
\caption{Radial velocities of stars.}
\begin{tabular}{lccccl}
\hline
\hline
ID$^1$ & R.A. (J2000) & Dec. (J2000) & $V_{\mathrm{LSR}}$ & $\epsilon V_{\mathrm{LSR}}$ & Remark$^2$ \\
  & (h:m:s) & ($^{\circ}:^{\prime}:^{\prime\prime}$) & (km s$^{-1}$) & (km s$^{-1}$) & \\
\hline
{\it BOES} & & & & & \\
\hline
  $-1$ & 05:22:46.54 & $+$33:25:11.2 &$  -2.2$ &  3.0 & HD 242935 \\
  $-2$ & 05:22:44.00 & $+$33:26:26.7 &$  -1.9$ &  8.6 & BD +33 1025 \\
  $-6$ & 05:22:29.30 & $+$33:30:50.4 &$  -3.8$ &  0.8 & HD 242908 \\
\hline
\it{ Hectochelle}$^3$ & & & & & \\
\hline
 $-18$ & 05:23:12.84 & $+$33:18:55.2 &$   2.0 $&  1.4 &   \\
 129 & 05:22:05.72 & $+$33:24:47.7 &$  15.7 $&  1.4 &   \\
 411 & 05:22:11.79 & $+$33:34:05.7 &$ -38.0 $&  0.3 &   \\
\hline
\label{tab3}
\end{tabular}
\end{minipage}
\begin{tabular}{@{}l@{}}
$^1$ Star IDs are taken from \citet{LSK14}. \\
$^2$ SB2: Double-lined spectroscopic binary, SB2?: Double-lined spectroscopic binary candidate.\\
$^3$ The radial velocities of stars with r-statistics larger than 6.\\
The full table is available electronically.
\end{tabular}
\end{table*}






\bsp	
\label{lastpage}
\end{document}